\documentclass[a4paper,11pt]{article}
\usepackage{pos}

\newcommand{\be}{\begin{equation}}
\newcommand{\ee}{\end{equation}}
\newcommand{\bea}{\begin{eqnarray}}
\newcommand{\eea}{\end{eqnarray}}

\def\gsim{\,\lower.25ex\hbox{$\scriptstyle\sim$}\kern-1.30ex%
\raise 0.55ex\hbox{$\scriptstyle >$}\,}
\def\lsim{\,\lower.25ex\hbox{$\scriptstyle\sim$}\kern-1.30ex%
\raise 0.55ex\hbox{$\scriptstyle <$}\,}

\newcommand{\gev}{\,\mbox{GeV}}

\newcommand{\qsq}{\ensuremath{Q^2} }
\newcommand{\gevsq}{\ensuremath{\mathrm{GeV}^2} }

\title{HERA data on azimuthal decorrelation and charged particle multiplicity spectra probing QCD dynamics and quantum entanglement effects} 
\ShortTitle{QCD dynamics and quantum entanglement}

\author*[a]{Zhoudunming Tu, for the H1 and ZEUS Collaborations}

\affiliation[a]{Department of Physics, Brookhaven National Laboratory, Upton, NY 11973, USA}

\emailAdd{zhoudunming@bnl.gov}

\abstract{
The azimuthal decorrelation angle between the leading jet and scattered lepton in deep inelastic scattering is studied with the ZEUS detector at HERA. The data was taken in the HERA II data-taking period and corresponds to an integrated luminosity of 330 $\rm{pb^{-1}}$. Azimuthal angular decorrelation has been proposed to study the $Q^{2}$ dependence of the evolution of the transverse momentum distributions (TMDs) and understand the small-$x$ region, providing unique insight to nucleon structure. Previous decorrelation measurements of two jets have been performed in proton-proton collisions at very high transverse momentum; these measurements are well described by perturbative QCD at next-to-leading order. The azimuthal decorrelation angle obtained in these studies shows good agreement with predictions from Monte Carlo models including leading order matrix elements and parton showers. 
\\

New experimental data on charged particle multiplicity distributions are presented, covering the kinematic ranges in momentum transfer $5<Q^{2}<100~\rm GeV^{2}$ and inelasticity $0.0375<y<0.6$. The data was recorded with the H1 experiment at the HERA collider in positron-proton collisions at a centre-of-mass energy of 320 GeV.  Charged particles are counted with transverse momenta larger than 150 MeV and pseudorapidity $-1.6<\eta_{\rm lab}<1.6$ in the laboratory frame, corresponding to high acceptance in the current hemisphere of the hadronic centre-of-mass frame. Charged particle multiplicities are reported on a two-dimensional grid of $Q^{2}$, $y$ and on a three-dimensional grid of $Q^{2}$, $y$ and $\eta$.
The observable is the probability $P(N)$ to observe $N$ particles in the given $\eta$ region. The data are confronted with predictions from Monte Carlo generators, and with a simplistic model based on quantum entanglement and strict parton-hadron duality.
}

\FullConference{%
  40th International Conference on High Energy physics - ICHEP2020\\
  July 28 - August 6, 2020\\
  Prague, Czech Republic (virtual meeting)
}

\begin{document}
\maketitle

\section{Introduction}
In proton-proton ($pp$) collisions, it is suggested that the decorrelation measurement of two high transverse momentum back-to-back jets is sensitive to perturbative QCD at next-to-leading order. It has been recently proposed to look at the decorrelation between the scattered lepton and the leading jet in electron-proton ($ep$) deep inelastic scattering (DIS) process. The azimuthal angular decorrelation can be used to probe the $Q^{2}$ evolution of the Transverse Momentum Distributions (TMDs) and understand the small-$x$~region. 

In the parton model~\cite{Bjorken:1969ja,Feynman:1969wa,Gribov:1968fc} formulated by Bjorken, Feynman, and Gribov, the bounded quarks and gluons of a nucleon are viewed as ``quasi-free" particles by an external hard probe in the infinite momentum frame. The parton that participates in the hard interaction with the probe, e.g., the virtual photon, is expected to be causally disconnected from the rest of the nucleon. On the other hand, the parton and the rest of the nucleon have to form a colour-singlet state due to colour confinement. In order to further understand the role of colour confinement in high energy collisions, it has been suggested~\cite{Klebanov:2007ws,Kharzeev:2017qzs} that the quantum entanglement of partons could be an important probe of the underlying mechanism of confinement. 

In recent years, the idea of considering quantum entanglement in high energy collisions has been realized and many interesting results have been found both theoretically and experimentally. For example, in a study by Tu et al.~\cite{Tu:2019ouv} based on data at the Large Hadron Collider (LHC), the entropy of charged particles produced in $pp$ collisions is found to have a strong correlation to the entanglement entropy predicted by the gluon density~\cite{Kharzeev:2017qzs}, which shows a first indication of quantum entanglement of partons inside the proton. However, in high energy $pp$ collisions, there are other phenomena that might play an important role in particle production, e.g., Multiple Parton Interaction (MPI), Colour Reconnection (CR), etc. Therefore, the entanglement of partons can be investigated in $ep$ DIS events with better-defined theoretical interpretations. 

In high energy $ep$ DIS, the hard interaction between the virtual photon and the parton defines a transverse spatial domain by a size of $1/Q$ within the target proton, where $Q$ is defined by the virtuality of the photon. The collision separates the target proton into a probed region and a proton remnant, denoted by region $A$ and $B$, respectively. In the parton model where collinear factorization is assumed, region $A$ and $B$ are expected to be causally disconnected and therefore have no correlation. However, if partons in region $A$ and $B$ are entangled quantum mechanically, the entanglement entropy of $A$ and $B$ would be identical, e.g., $S_{\rm A}=S_{\rm B}$. Based on Refs.~\cite{Kharzeev:2017qzs,Tu:2019ouv}, the entanglement entropy in DIS was found to have a simple relation with the gluon density $xG(x,Q^{2})$ in the low-$x$~limit as, $S_{\rm parton}=\ln{[xG(x)]}$\footnote{Hereafter the $Q^{2}$ dependence of gluon density is dropped for simplicity, denoted as $xG(x)$}. This was inspired by a well known result for the entanglement entropy in $(1+1)$ conformal field theory, where the length of the studied region in the context of DIS is $(1/mx)$\footnote{In the target rest frame, $m$ is the proton rest mass, $(1/mx)\sim(1/x)$} which is closely related to parton distributions.
In addition, it is suggested~\cite{Kharzeev:2017qzs} that the proportionality is expected to be valid between the final-state hadron entropy, $S_{\rm hadron}$, and the initial-state parton entropy, $S_{\rm parton}$, due to the ``parton liberation"~\cite{Mueller:1999fp} and "local parton-hadron duality (LPHD)"~\cite{Dokshitzer_1991} pictures. Therefore, the entanglement entropy $S_{\rm A}$ (equivalent to notation $S_{\rm EE}$) can be revealed by the final-state hadron entropy, e.g., 
\be\label{eq1}
S_{\rm parton} = \ln{[xG(x)]} = S_{\rm hadron} = -\sum{P(N)\ln{P(N)}},
\ee
\noindent where $P(N)$ is the charged particle multiplicity distribution.

Despite the new idea of relating final-state hadron multiplicity to the entanglement entropy of partons, charged particle production has been extensively studied in high energy collisions over many decades, from electron-positron ($e^{+}e^{-}$) scattering to heavy ion collisions. For reviews, see Ref.~\cite{Becattini1995} and the references therein. On the one hand, the exact particle production mechanism and quantitative prediction of multiplicity distributions are not yet completely understood in hadron (nucleus) collider experiments due to the complicated substructure of nucleon and parton fragmentation. For example, no first-principle calculation can describe the multiplicity distributions at the LHC in $pp$ collisions, and no phenomenology model can reproduce those distributions without significant tuning~\cite{Corke:2010yf}. On the other hand, the measurement of entanglement entropy of partons via final-state hadrons might provide a new perspective to particle productions without directly considering fragmentation. For instance, the entanglement entropy in high energy collisions implies a natural upper limit on the particle multiplicity density~\cite{Kharzeev:2017qzs}, similar to the prediction from the theory of Color Glass Condense with gluon saturation~\cite{Gelis:2010nm}.

\section{Results}

\begin{figure}[thb]
\includegraphics[width=4.50in]{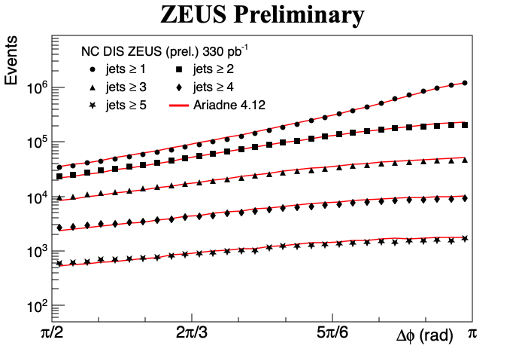}
  \caption{ \label{fig:figure_0} Angular decorrelation in $ep$ DIS at HERA using the ZEUS detector. }
\end{figure}

In Fig.~\ref{fig:figure_0}, the angular decorrelation between the scattered lepton and the leading jet in $ep$ DIS at HERA using the ZEUS detector is shown. The data is shown in different categories, where different number of jets per event is compared. It is found that the angular distribution broadens if the number of jets per event increases. The MC predictions based on Ariadne describe the data well. 

\begin{figure}[thb]
\includegraphics[width=5.50in]{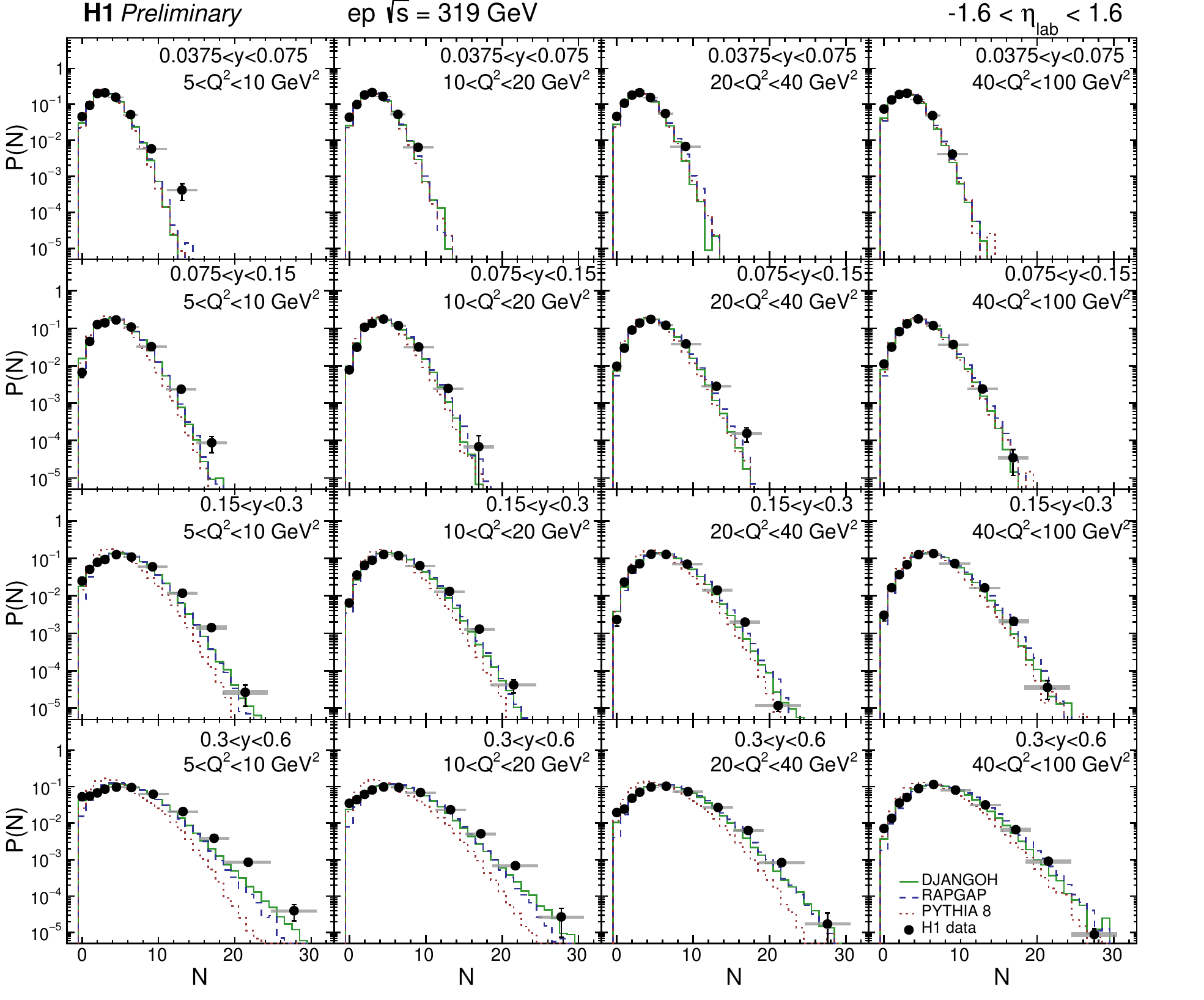}
  \caption{ \label{fig:figure_1} Multiplicity $P(N)$ distributions in $ep$ DIS at $\sqrt{s}=319$\gev~between $|\eta_{_{\rm{lab}}}|<1.6$ in the lab frame. }
\end{figure}

The charged particle multiplicity distributions in $ep$ DIS at $\sqrt{s}=319$\gev~are measured between $|\eta_{_{\rm{lab}}}|<1.6$ in the lab frame, shown in Fig.~\ref{fig:figure_1}. Different \qsq and $y$ bins are shown in different panels, where the \qsq ranges between 5 to 100 $\gevsq$ and $y$ is between 0.0375 to 0.6. The $P(N)$ distributions are fully unfolded, where the statistical uncertainty is denoted by the error bar and the systematic uncertainty is represented by the shaded box. The data are compared with generated truth level of the MC generators of DJANGOH, RAPGAP, and PYTHIA 8. 

\begin{figure}[thb]
\includegraphics[width=2.80in]{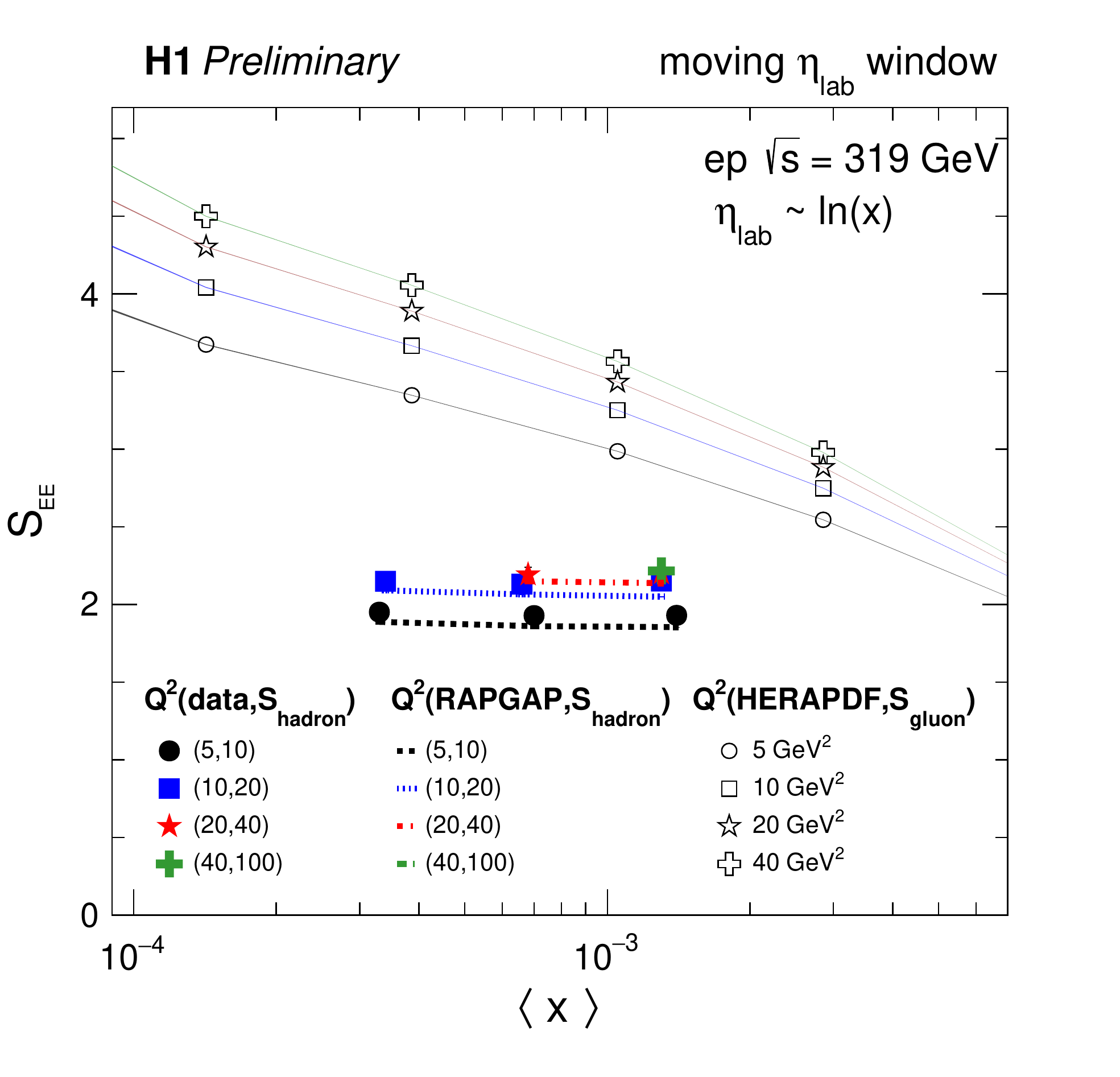}
\includegraphics[width=2.80in]{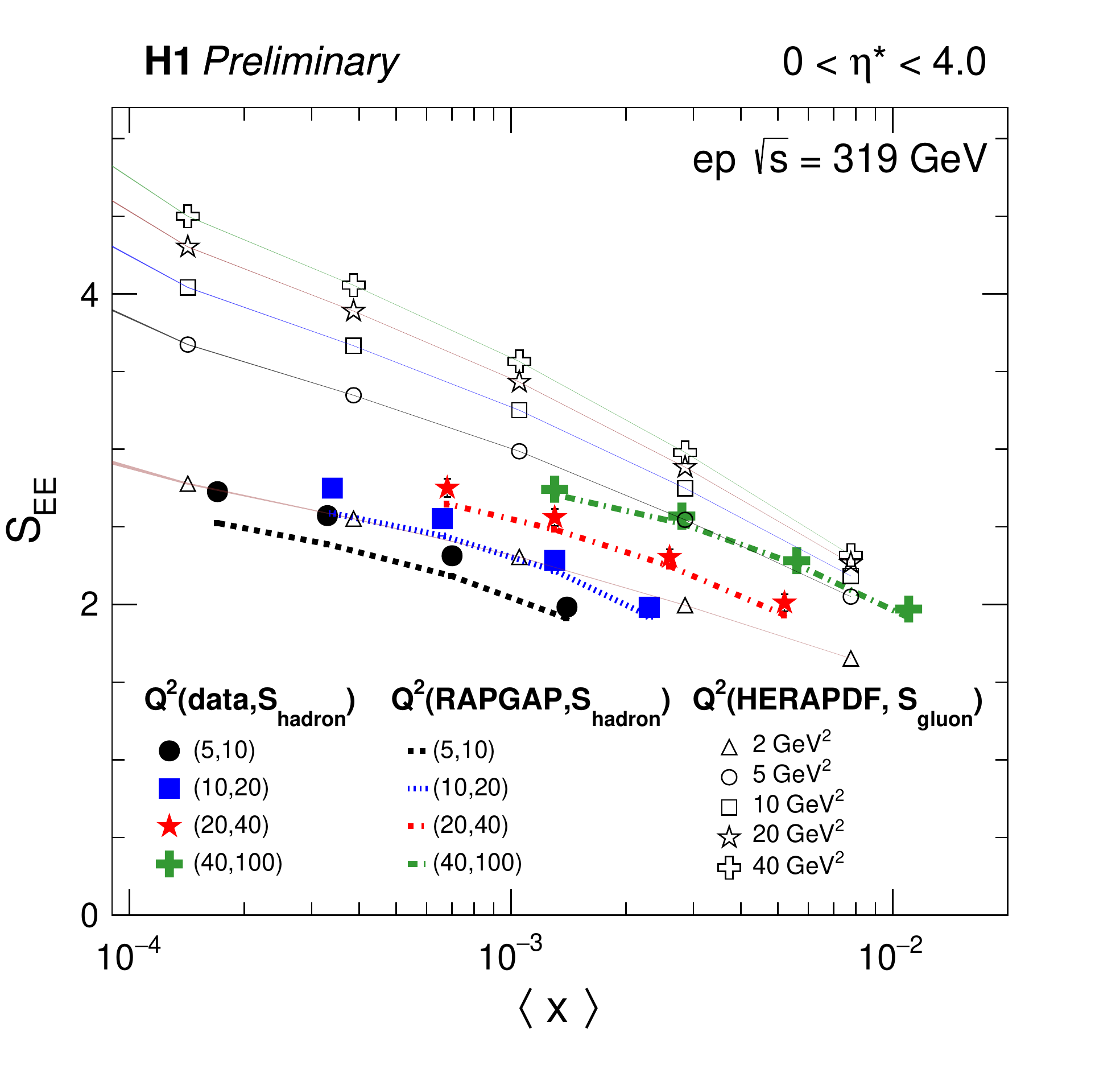}
  \caption{ \label{fig:figure_3} Hadron entropy measured in $ep$ DIS at $\sqrt{s}=319$\gev~between $|\eta_{_{\rm{lab}}}|<1.6$ in the lab frame (left) and $0<\eta^{*}<4.0$ in the HCM frame (right).}
\end{figure}

It is recently suggested by Refs.~\cite{Kharzeev:2017qzs,Tu:2019ouv} that the hadron entropy of final-state particles can be calculated based on the charged particle multiplicity distributions, which might indicate a deep connection to the entanglement entropy of gluons at low-$x$. In Fig.~\ref{fig:figure_3} left, the entropy of final-state hadron, $S_{\rm hadron}$, is studied as a function of $\langle $x$\rangle$ in different \qsq bins. The total uncertainty is indicated by the error bar. For each different $\langle x \rangle$ (or $y$) bin, the selected pseudorapidity window in the lab frame is used for measuring the multiplicity, e.g., $-1.2<\eta_{_{\rm{lab}}}<0.2$ at $\langle x \rangle\sim3\times10^{-4}$, $-0.5<\eta_{_{\rm{lab}}}<0.9$ at $\langle x \rangle\sim7\times10^{-4}$, and $-0.2<\eta_{_{\rm{lab}}}<1.6$ at $\langle x \rangle\sim1.3\times10^{-3}$. Similar to the observable studied in Ref.~\cite{Tu:2019ouv}, the varying $\eta_{_{\rm{lab}}}$ range is to approximately match the rapidity of the scattered quark from the DIS process in a leading order picture. The same observable is studied using MC event generator RAPGAP, which qualitatively agrees with the data at each measured \qsq bin. On the other hand, the predictions from entanglement entropy based on the gluon density $xG(x)$ are also shown for comparison at various of \qsq values, indicated by the open markers with coloured bands. The prediction and the data are found to be inconsistent. 

Taking one step further, it is possible to measure the hadron entropy of particles from the current fragmentation hemisphere with 4 units of pseudorapidity coverage, shown in Fig.~\ref{fig:figure_3} right. The hadron entropy based on multiplicity distributions are studied as a function of $\langle x \rangle$ in different \qsq bins within a fixed pseudorapidity range $0<\eta^{*}<4.0$ in the hadronic center-of-mass (HCM) frame. The MC generally describes the data well except for low \qsq at low-$x$, while the entanglement entropy based on gluon density fails to describe the data. 

\section{Summary}

The angular decorrelation between the scattered lepton and the leading jet has been investigated in $ep$ DIS using the ZEUS detector at $\sqrt{s}=319$\gev. The MC prediction can generally describe the data well. The charged particle multiplicity distributions, $P(N)$, in deep inelastic scattering events at $\sqrt{s}=319$\gev~using the H1 detector at HERA are measured. The hadron entropy based on multiplicity distributions are found to be inconsistent with the prediction from entanglement entropy of gluons, while further theoretical calculations of entanglement entropy with \qsq evolution including sea partons is needed for a proper comparison to the measured data.

\bibliographystyle{aps}
\bibliography{reference}

\begin{thebibliography}{10}
\providecommand{\url}[1]{\texttt{#1}}
\providecommand{\urlprefix}{URL }
\providecommand{\eprint}[2][]{\url{#2}}

\bibitem{Bjorken:1969ja}
J.~D. Bjorken and E.~A. Paschos, {Inelastic Electron Proton and gamma Proton
  Scattering, and the Structure of the Nucleon}, Phys. Rev. \textbf{185}, 1975
  (1969).

\bibitem{Feynman:1969wa}
R.~P. Feynman, {The behavior of hadron collisions at extreme energies}, Conf.
  Proc. \textbf{C690905}, 237 (1969).

\bibitem{Gribov:1968fc}
V.~N. Gribov, {A reggeon diagram technique}, Sov. Phys. JETP \textbf{26}, 414
  (1968), [Zh. Eksp. Teor. Fiz.53,654(1967)].

\bibitem{Klebanov:2007ws}
I.~R. Klebanov, D.~Kutasov, and A.~Murugan, {Entanglement as a probe of
  confinement}, Nucl. Phys. \textbf{B796}, 274 (2008), \eprint{0709.2140}.

\bibitem{Kharzeev:2017qzs}
D.~E. Kharzeev and E.~M. Levin, {Deep inelastic scattering as a probe of
  entanglement}, Phys. Rev. \textbf{D95}, 114008 (2017), \eprint{1702.03489}.

\bibitem{Tu:2019ouv}
Z.~Tu, D.~E. Kharzeev, and T.~Ullrich, {The EPR paradox and quantum
  entanglement at sub-nucleonic scales}  (2019), \eprint{1904.11974}.

\bibitem{Mueller:1999fp}
A.~H. Mueller, {Toward equilibration in the early stages after a high-energy
  heavy ion collision}, Nucl. Phys. \textbf{B572}, 227 (2000),
  \eprint{hep-ph/9906322}.

\bibitem{Dokshitzer_1991}
Y.~L. Dokshitzer, V.~A. Khoze, and S.~I. Troyan, On the concept of local
  parton-hadron duality, Journal of Physics G: Nuclear and Particle Physics
  \textbf{17}, 1585 (1991).

\bibitem{Becattini1995}
F.~Becattini, A thermodynamical approach to hadron production in ee collisions,
  Zeitschrift f{\"u}r Physik C Particles and Fields \textbf{69}, 485 (1995).

\bibitem{Corke:2010yf}
R.~Corke and T.~Sjostrand, {Interleaved Parton Showers and Tuning Prospects},
  JHEP \textbf{03}, 032 (2011), \eprint{1011.1759}.

\bibitem{Gelis:2010nm}
F.~Gelis, E.~Iancu, J.~Jalilian-Marian, and R.~Venugopalan, {The Color Glass
  Condensate}, Ann. Rev. Nucl. Part. Sci. \textbf{60}, 463 (2010),
  \eprint{1002.0333}.

\end{thebibliography}

\end{document}